
\magnification=1200
\baselineskip=10pt

\rightline{UR-1332\ \ \ \ \ \ \ }
\rightline{ER-40685-782}

\bigskip

\centerline{\bf DAVEY-STEWARTSON EQUATION FROM A ZERO CURVATURE}
\medskip
\centerline{\bf AND A SELF-DUALITY CONDITION}

\vskip .75in

\centerline{J. C. Brunelli}

\medskip

\centerline{and}

\medskip

\centerline{Ashok Das}

\medskip

\centerline{Department of Physics and Astronomy}
\centerline{University of Rochester}
\centerline{Rochester, NY 14627}

\vskip 1in

\baselineskip=20pt

\centerline{\bf \underbar{Abstract}}

\medskip

We derive the two equations of Davey-Stewartson type from a zero curvature
condition associated with SL(2,{\bf R}) in $2+1$ dimensions.
We show in general how a $2+1$ dimensional zero curvature condition can be
obtained from the self-duality condition in $3+3$ dimensions and show in
particular how the Davey-Stewartson equations can be obtained from the
self-duality condition associated with SL(2,{\bf R}) in $3+3$ dimensions.

\vfill\eject

\noindent {\bf I. Introduction:}

\medskip

Integrable field theories are quite interesting and a lot of their
properties are well studied in $1+1$ dimensions [1,2].  In $2+1$ dimensions,
however, only a few are known and very little is known about their
structure.  While the KP equation (Kadomtsev-Petviashvilli equation)
 [3] has
been the subject of intense studies in recent years, the Davey-Stewartson
equations [4] -- which are also interesting in their own right -- have not
received as much attention.  In particular, even though one knows the
matrix Lax pair for this equation [5], a scalar Lax formulation or even a zero
curvature formulation [2,6] for the system does not exist.  Similarly, the
Gelfand-Dikii bracket structures [7] for such systems are also not understood.
In fact, it is not known how most of the interesting properties of $1+1$
dimensional theories extend to higher space-time dimensions.  In this
letter, we make a modest contribution to such studies.  In particular, we
show, in sec. II how the Davey-Stewartson equation can be obtained from a
zero curvature condition [2,6]
 associated with SL(2,{\bf R}) in $2+1$ dimensions.
 In sec. III, we further derive these equations from the self-duality
conditions [8-10]
 associated with SL(2,{\bf R}) in $3+3$ dimensions.  We end with
a brief conclusion in sec. IV.

\medskip

\noindent {\bf II. Zero Curvature Formulation:}

\medskip

There are two Davey-Stewartson equations [4,5] in $2+1$ dimensions commonly
known as DSI and DSII equations and both are known to be integrable.  DSI
has the form
$$\eqalign{&i \dot q = - {1 \over 2} \big( \partial^2_1 +
 \partial^2_2 \big) q - \big( qr + \partial_2 \phi \big) q\cr
&\big( \partial^2_1 - \partial^2_2 \big) \phi - 2
\partial_2 (qr) =0 \cr} \eqno(1)$$
where $q$ and $\phi$ are the basic dynamical variables.  $q$ is complex
while $\phi$ is real and $r= \pm q^*$.  In our notation $\partial_0,\
\partial_1$ and $\partial_2$ represent derivatives with respect to $t, \ x,
\ y$ respectively.  The other equation, DSII, on the other hand, has the
structure
$$\eqalign{&i \dot q = - {1 \over 2} \big( \partial^2_1 -
 \partial^2_2 \big) q + \big( qq^* + \partial_2 \phi \big) q\cr
&\big( \partial^2_1 + \partial^2_2 \big) \phi + 2
\partial_2 (qq^*) =0 \cr} \eqno(2)$$
We note here that Eq. (2) can be obtained from Eq. (1) by letting
$y \rightarrow - iy$, $\phi \rightarrow i \phi$ and identifying $r=-q^*$.
Similarly, we can obtain Eq. (1) from Eq. (2) by letting $y \rightarrow
-iy$, $\phi \rightarrow i \phi$ and identifying $q^* = -r$.
Since DSI and DSII can be obtained from each other through simple
redefinitions, we will concentrate, for simplicity, on only one of them,
namely, DSII given in Eq. (2).

The nonlinear equations (2) can also be written in terms of linear
equations of the form [5]
$$\pmatrix{\partial_1 + i \partial_2 &-q\cr
\noalign{\vskip 4pt}
-q^* &\partial_1 - i \partial_2\cr}
\pmatrix{\psi_1 \cr
\noalign{\vskip 4pt}
\psi_2\cr} = 0
\eqno(3)$$

\medskip

$$i \partial_0 \pmatrix{\psi_1\cr
\noalign{\vskip 5pt}%
\psi_2\cr}
= \pmatrix{\partial^2_2 + {i \over 2} \big( (\partial_1 - i \partial_2)
 \phi \big) + {1 \over 2} q^*q
&i q \partial_2 - {1 \over 2} \big( (\partial_1 - i \partial_2) q \big)\cr
\noalign{\vskip 6pt}%
i q^* \partial_2 + {1 \over 2} \big( (\partial_1 + i \partial_2) q^*
\big)
&-\partial^2_2 + {i \over 2} \big( (\partial_1 + i \partial_2 )
\phi \big) - {1 \over 2} q^*q\cr}
\pmatrix{\psi_1\cr
\psi_2\cr} \eqno(4)$$

\smallskip

\noindent The consistency
 of Eqs. (3) and (4), then give rise to the Davey-Stewartson
equations (2).  Eqs. (3) and (4), therefore, define the Lax pair
for the system, but note that these Lax operators
$$L = \pmatrix{\partial_1 + i \partial_2 &-q\cr
\noalign{\vskip 4pt}%
-q^* &\partial_1 - i \partial_2\cr}  \eqno(5)$$

\medskip

$$B
= \pmatrix{\partial^2_2 + {i \over 2} \big( (\partial_1 - i \partial_2)
 \phi \big) + {1 \over 2} q^*q
&i q \partial_2 - {1 \over 2} \big( (\partial_1 - i \partial_2) q
 \big) \cr
\noalign{\vskip 6pt}%
i q^* \partial_2 + {1 \over 2} \big( (\partial_1 + i \partial_2) q^*
 \big)
&-\partial^2_2 + {i \over 2} \big( (\partial_1 + i \partial_2 )
\phi \big) - {1 \over 2} q^*q\cr} \eqno(6)$$

\smallskip

\noindent unlike the $1+1$
 dimensional case are matrix Lax operators.  Furthermore,
even though these operators are matrices, the consistency condition is not
equivalent to a zero curvature condition.

To obtain a zero curvature formulation
 [2,6] of the Davey-Stewartson equations,
let us note that these equations can be thought of as a higher dimensional
generalization of the nonlinear Schr\"odinger equation in $1+1$ dimension.
The nonlinear Schr\"odinger equation, on the other hand, is known to
arise from a zero curvature condition associated with SL(2,{\bf R}) in $1+1$
dimensions [2].  Thus, let us choose SL(2,{\bf R}) as the appropriate group for
our potentials.  With the explicit representations of the generators
$$\sigma_+ = \pmatrix{0 &1\cr
\noalign{\vskip 4pt}
0 &0\cr} \qquad \sigma_- = \pmatrix{0 &0\cr
\noalign{\vskip 4pt}1 &0\cr}
\qquad
\sigma_3 = \pmatrix{1 &0\cr
\noalign{\vskip 4pt}
0 &-1} \eqno(7)$$
let us choose
$$\eqalign{A_0 &= - {i  \over 2} \big( qq^* + \partial_2 \phi +
 3 \lambda^2 \big) \sigma_3 - 3 \lambda  q^*
\sigma_+ - 3 \lambda q \sigma_-\cr
A_1 &= \sqrt{2}\ i \lambda  \sigma_3 + {1 \over \sqrt{2}}\
q^* \sigma_+ + {1
\over \sqrt{2}}\ q \sigma_-\cr
A_2 &= - {i \lambda  \over \sqrt{2}}\ \sigma_3 +
{1 \over \sqrt{2}}\ q^* \sigma_+ +
{1 \over \sqrt{2}} \ q \sigma_-\cr}\eqno(8)$$
Here $\lambda$ is a constant parameter.  The curvatures (field strengths)
associated with these potentials can be easily constructed from
$$F_{\mu \nu} = \partial_\mu A_\nu - \partial_\nu A_\mu
+ \left[ A_\mu , A_\nu \right] \eqno(9)$$
Requiring the curvatures to vanish, we expect nine equations.
$F_{01} = 0$ gives
$$\eqalignno{&\partial_1 \big( qq^* + \partial_2 \phi \big) = 0 &(10)\cr
&\dot q + 3 \sqrt{2}\ \lambda \partial_1 q + i  \big( qq^*
+ \partial_2 \phi - 9 \lambda^2 \big) q = 0 &(11)\cr
&\dot q^* + 3 \sqrt{2}\ \lambda \partial_1 q^* - i
\big( qq^* + \partial_2 \phi - 9 \lambda^2 \big) q^* = 0 &(12)\cr}$$
Similarly, $F_{02} = 0$ gives
$$\eqalignno{&\partial_2 \big( qq^* + \partial_2 \phi \big) = 0 &(13)\cr
&\dot q + 3 \sqrt{2}\ \lambda   \partial_2 q + i  \big( qq^*
+ \partial_2 \phi + 9 \lambda^2 \big) q = 0 &(14)\cr
&\dot q^* + 3 \sqrt{2}\ \lambda  \partial_2 q^* - i
\big( qq^* + \partial_2 \phi + 9 \lambda^2 \big) q^* = 0 &(15)\cr}$$
Finally, $F_{12}= 0$ leads to only two equations
$$\eqalignno{&\big( \partial_1 -  \partial_2 \big) q - 3 \sqrt{2}\
i \lambda  q = 0 &(16)\cr
&\big( \partial_1 -  \partial_2 \big) q^* + 3 \sqrt{2}\
i \lambda  q^* = 0 &(17)\cr}$$

We note from Eq. (16) that
$$3 \sqrt{2}\ i \lambda  \left( \partial_1 +
\partial_2 \right) q = \left( \partial^2_1 -
\partial^2_2 \right) q \eqno(18)$$
Adding Eqs. (11) and (14) and using Eq. (18) we obtain
$$\eqalign{&\dot q + {3 \over \sqrt{2}}\ \lambda \big( \partial_1 +
  \partial_2 \big) q + i  \big( qq^* + \partial_2 \phi \big)q
 = 0\cr
{\rm or}, \qquad &i \dot q + {1 \over 2}\ \big( \partial_1^2 -
  \partial_2^2 \big) q -
  \big( qq^* + \partial_2 \phi \big)q
 = 0\cr}\eqno(19)$$
which is, of course, the same as the first of
 Eq. (2).  Similarly, from Eqs. (10),
(13), and (17) we obtain
$$\partial_2 \left[ \left( \partial^2_1 +  \partial^2_2 \right)
\phi + 2  \partial_2 ( qq^*) \right] = 0 \eqno(20)$$
which with a suitable field redefinition of $\phi$ can be written as
$$\left( \partial^2_1 +  \partial^2_2 \right) \phi + 2
\partial_2 (qq^*) = 0 \eqno(21)$$
This is the second part  of the Davey-Stewartson equations (Eq. (2)).  Thus,
we see that both the Davey-Stewartson equations can be derived from a zero
curvature condition associated with SL(2,{\bf R}) in $2+1$ dimensions.

\medskip

\noindent {\bf III. Self-Dual Formulation:}

\medskip

It is well known that the $1+1$ dimensional integrable models can be
obtained from a self-duality condition associated with Yang-Mills field
strengths belonging to SL(2,{\bf R}) in $2+2$ dimensions [8-10].
  The KP equation
is also known to result from a self-duality condition on Yang-Mills field
strengths belonging to SL(2,{\bf R}) in $3+3$ dimensions [11]. Since
Davey-Stewartson equation is a $2+1$ dimensional equation like the KP
equation, we examine the self-duality conditions in $3+3$ dimensions.  The
self-duality conditions in dimensions higher than four are quite tricky and
we refer the interested readers to ref. 11 for details.  Here we only note
that the self-duality conditions on the field strengths in $3+3$ dimensions
lead to
$$\eqalign{F^\prime_{02} &= 0\cr
F^\prime_{05} &= 0\cr
F^\prime_{25} &= 0\cr
F^\prime_{13} &= 0\cr
F^\prime_{14} &= 0\cr
F^\prime_{34} &= 0\cr
F^\prime_{01} &+ F^\prime_{23} + F^\prime_{54} = 0\cr}\eqno(22)$$
Here
$$F^\prime_{\mu \nu} = \partial_\mu A^\prime_\nu - \partial_\nu
A^\prime_\mu +
\left[ A^\prime_\mu , A^\prime_\nu \right]
\qquad \mu , \nu = 0 , 1, \dots , 5 \eqno(23)$$
are the $3+3$ dimensional field strength tensors
(We use primes to emphasize that these are $3+3$ dimensional objects.).
 Let us next choose
$x^0 = t,\ x^2 = x,\ x^5 = y$ and if we identify
$$\eqalign{A^\prime_0 &= A_0 = - {i  \over 2}\ \big(
qq^* + \partial_2 \phi + 3 \lambda^2 \big) \sigma_3 -
3 \lambda  q^* \sigma_+ - 3 \lambda  q \sigma_-\cr
A^\prime_2 &= A_1 = \sqrt{2}\ i \lambda  \sigma_3 + {1 \over
\sqrt{2}}\ q^* \sigma_+ + {1 \over \sqrt{2}}\ q \sigma_-\cr
A^\prime_5 &= A_2 = - {i \lambda \over \sqrt{2}}\  \sigma_3 +
{1 \over \sqrt{2}}\ q^* \sigma_+ + {1 \over \sqrt{2}} \ q \sigma_-\cr}
\eqno(24)$$
then the first three equations of (22) would merely give the
Davey-Stewartson equation of Eq. (2) as we have seen in the previous
section.  The rest of the equations can be solved by
noting that with the reduction conditions (This is different from ref. 11.)
$$\eqalign{&\partial_0 - \partial_1 = 0 = \partial_0 -
\partial_4\cr
&\partial_2 - \partial_3 = 0\cr}\eqno(25)$$
and with the identification
$$A^\prime_1 = A^\prime_4 = A^\prime_0 \qquad {\rm and}
\qquad A^\prime_3 = A^\prime_2 \eqno(26)$$
we have
$$\eqalign{F^\prime_{13} &= \partial_1 A^\prime_3 - \partial_3 A^\prime_1
 + \big[ A^\prime_1 , A^\prime_3 \big] = \partial_0 A^\prime_2 -
 \partial_2 A^\prime_0 + \big[ A^\prime_0 , A^\prime_2 \big] =
F^\prime_{02} = 0\cr
F^\prime_{14} &= \partial_1 A^\prime_4 - \partial_4 A^\prime_1
 + \big[ A^\prime_1 , A^\prime_4 \big] = \partial_0 A^\prime_0 -
 \partial_0 A^\prime_0 + \big[ A^\prime_0 , A^\prime_0 \big] = 0\cr
F^\prime_{34} &= \partial_3 A^\prime_4 - \partial_4 A^\prime_3 +
  \big[ A^\prime_3 , A^\prime_4 \big] = \partial_2 A^\prime_0 -
 \partial_0 A^\prime_2 + \big[ A^\prime_2 , A^\prime_0 \big] =
- F^\prime_{02} = 0\cr
F^\prime_{01} &= \partial_0 A^\prime_1 - \partial_1 A^\prime_0
 + \big[ A^\prime_0 , A^\prime_1 \big] = \partial_0 A^\prime_0 -
 \partial_0 A^\prime_0 + \big[ A^\prime_0 , A^\prime_0 \big] = 0\cr
F^\prime_{23} &= \partial_2 A^\prime_3 - \partial_3 A^\prime_2
 + \big[ A^\prime_2 , A^\prime_3 \big] = \partial_2 A^\prime_2 -
 \partial_2 A^\prime_2 + \big[ A^\prime_2 , A^\prime_2 \big] = 0\cr
F^\prime_{54} &= \partial_5 A^\prime_4 - \partial_4 A^\prime_5
 + \big[ A^\prime_5 , A^\prime_4 \big] = \partial_5 A^\prime_0 -
 \partial_0 A^\prime_5 + \big[ A^\prime_5 , A^\prime_0 \big] =
- F^\prime_{05} = 0\cr}\eqno(27)$$
Thus, we see that the rest of the equations in (22) are automatically
satisfied with our choice of reduction by virtue of the first three
equations which with the ansatz of Eq. (24) lead to the Davey-Stewartson
equation.  This shows how the Davey-Stewartson equations can be
obtained from
 the self-duality condition on Yang-Mills field strengths belonging to
SL(2,{\bf R}) in $3+3$ dimensions.  This result, in fact, is quite general
and shows that any $2+1$ dimensional equation which can be formulated as a
zero curvature condition, can also be obtained from a self-duality
condition on Yang-Mills fields in $3+3$ dimensions with our choice of
reduction and gauge potential identifcations given in Eqs. (25) and (26).
This result is, in fact, quite similar to the general procedure outlined
for self-duality in $2+2$ dimensions [12].
This, therefore, gives an alternate to the derivation of the KP
equation from a self-duality condition proposed in ref. 11 also.

\medskip

\noindent {\bf IV. Conclusion:}

\medskip

We have shown how the two Davey-Stewartson equations can be obtained from a
zero curvature condition associated with SL(2,{\bf R}) in $2+1$ dimensions.
 We have also shown how a general equation in $2+1$ dimensions
resulting from a zero curvature condition can be obtained from a self-duality
condition on Yang-Mills field strengths in $3+3$ dimension with appropriate
reduction.  In particular, we have shown that the Davey-Stewartson
equations can be obtained from a self-duality condition associated with
field strengths belonging to SL(2,{\bf R}).

\medskip

\noindent {\bf Acknowledgements:}

\medskip

This work was supported in part by the U.S. Department of Energy Grant no.

\noindent DE-FG-02-91ER40685 and by CNPq, Brazil.

\vfill\eject

\noindent {\bf References:}

\medskip

\item{1.} L D. Faddeev and L. A. Takhtajan, Hamiltonian Methods in the
Theory of Solitons (Springer, 1987).

\item{2.} A. Das, Integrable Models (World Scientific, 1989).

\item{3.} B. B. Kadomtsev and V. I. Petviashvili, Sov. Phys. Dokl.
{\bf 15}, 539 (1971).

\item{4.} A. Davey and K. Stewartson, Proc. R. Soc. {\bf A338}, 101 (1974).
\item{  } V. D.
 Djordjevic and L. G. Redekopp, J. Fluid Mech. {\bf 79}, 703 (1977).

\item{5.} A. S. Fokas and M. J. Ablowitz, J. Math. Phys. {\bf 25}, 2494
(1984); D. J. Kaup, Inverse Problems {\bf 9}, 417 (1993).

\item{6.} S. S. Chern and C. K. Peng, Manuscr. Math. {\bf 28}, 207 (1979);
M. Crampin, F. A. Pirani and D. C. Robinson, Lett. Math. Phys. {\bf 2}, 15
(1977).

\item{7.} I. M. Gelfand and L. A. Dickii, Funct. Anal. Appl. {\bf 10},
259 (1976); M. Adler, Invent. Math. {\bf 50}, 219 (1979); V. G. Drinfeld
and V. V. Sokolov, J. Sov. Math. {\bf 30}, 1975 (1985).

\item{8.} R. S. Ward, Nucl. Phys. {\bf B236}, 381 (1984); Philos. Trans. R.
Soc. London {\bf A315}, 451 (1985); also in Field Theory, Quantum Gravity
and Strings, ed. H. J. deVega and N. Sanchez, Vol. 2 p.106.

\item{9.} L. J. Mason and G. A. J. Sparling, Phys. Lett. {\bf A137}, 29
(1989).

\item{10.} I. Bakas and D. A. Depireux, Int. J. Mod. Phys. {\bf A7}, 1767
 (1992).

\item{11.} A. Das, Z. Khviengia and E. Sezgin, Phys. Lett. {\bf B289}, 347
(1992).

\item{12.} A. Das and C. A. P. Galv\~ao, Mod. Phys. Lett. {\bf A8}, 647
(1993);

\item{  } A. Das and C. A. P. Galv\~ao, Mod. Phys. Lett. {\bf A8}, 1399
 (1993).

\end